\documentclass[aps,pre,onecolumn]{revtex4}

\usepackage{graphicx,bm,hyperref,amssymb,amsmath}
\newcommand{\bi}{\begin{itemize}}
\newcommand{\ei}{\end{itemize}}
\newcommand{\ben}{\begin{enumerate}}
\newcommand{\een}{\end{enumerate}}
\newcommand{\be}{\begin{equation}}
\newcommand{\ee}{\end{equation}}
\newcommand{\bea}{\begin{eqnarray}} 
\newcommand{\eea}{\end{eqnarray}}
\newcommand{\bc}{\begin{center}}
\newcommand{\ec}{\end{center}}

\newcommand{\ie}{{\it i.e.\ }}
\newcommand{\eg}{{\it e.g.\ }}
\newcommand{\pd}[1]{\frac{\partial}{\partial #1}}
\newcommand{\pdd}[1]{\frac{\partial^2}{\partial #1^2}}

\newcommand{\mbf}[1]{{\bm #1}}
\newcommand{\etal}{{\it et.\ al.\ }}

\newcommand{\ino}{\int_\Omega}
\newcommand{\ke}{k}                     
\newcommand{\ph}{\phi}                  
\newcommand{\melp}{{MERSA}}                
\newcommand{\phr}{\ensuremath{\tilde{\ph}}}    

\newtheorem{pro}{Proposition}

\begin{document}
\title{
Analytic steady-state space use patterns
and rapid computations in mechanistic home range analysis}
%
%
\author{Alex H. Barnett}
\affiliation{Department of Mathematics, 6188 Kemeny Hall,
Dartmouth College, Hanover NH 03755, USA}
\email{ahb@math.dartmouth.edu}
\author{Paul R. Moorcroft}
\affiliation{OEB Dept, Harvard University, 22 Divinity Ave,
Cambridge MA 02138, USA}
\date{\today}

\begin{abstract}
Mechanistic home range models are important tools in modeling
animal dynamics in spatially-complex environments.
We introduce a class of 
stochastic models for animal movement in a habitat of varying preference.
Such models interpolate between spatially-implicit resource selection
analysis (RSA)
and advection-diffusion models, possessing these two models as
limiting cases. We find a closed-form solution
for the steady-state (equilibrium) probability distribution $u^*$
using a factorization of the redistribution operator
into symmetric and diagonal parts.
How space use is controlled by the preference function
$w$ then depends on the characteristic width of the redistribution kernel:
when $w$ changes rapidly compared to this width, $u^* \propto w$,
whereas on global scales large compared to this width, $u^* \propto w^2$.
We analyse the behavior at discontinuities in $w$
which occur at habitat type boundaries.
We simulate the dynamics of space use given
two-dimensional prey-availability data and explore the effect of
the redistribution kernel width.
Our factorization allows such numerical
simulations to be done extremely fast;
we expect this to aid the computationally-intensive
task of model parameter fitting and inverse modeling.
\end{abstract}
\maketitle

\begin{figure}[t]  
\bc\includegraphics[width=5in]{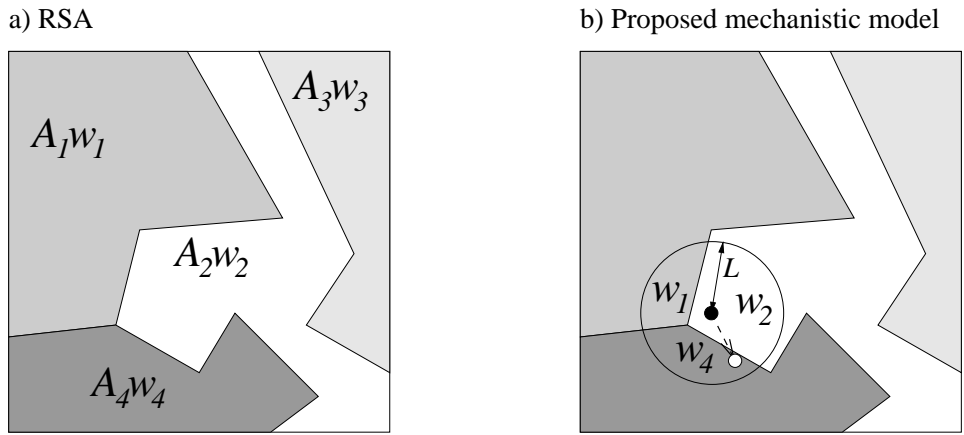}\ec
\caption{
\label{fig:bigpic}
Schematic comparing conventional resource selection analysis (RSA)
against our proposed model for space use, in two dimensions.
Shaded regions show areas of habitat in which the
preference function $w(x)$ is constant.
a) RSA. The probability of finding an animal in each region is proportional
to the product of preference (resource selection function)
$w_j$ and area of the region $A_j$.
There is no account taken of the animal's current habitat type or
location within the habitat.
b) Proposed \melp\ model for space use.
The animal responds only to the preference function
within a distance scale $L$ from its
current location (black dot). Each time step the future location
(white dot) is chosen randomly from a localized distribution
biased by the preference function.
}
\end{figure} 

\section{Introduction}



Due to the influences of habitats on the availability of food, shelter,
mates and risk of predation, patterns of animal space use are strongly
influenced by the spatial distribution of habitat types across
landscapes. A widespread technique for analyzing such relationships
between habitats
and patterns of animal
space use has been resource selection analysis (RSA)
\cite{johnsonrsa,manly,boycerev}, in which the intensity of space use
is assumed to reflect
an underlying resource selection function
giving an individual's
preference for the habitat type found at that location. However, implicit
in this approach
is an assumption that all habitats are
equally accessible to an individual regardless of its current
position: no account is made of the individual's finite
movement speed or the spatial geometry of habitats on the landscape
on which it moves \cite{mhra,recon}.

More recently, an alternative framework
has emerged in the form of mechanistic home range
models \cite{lewis,moorcroftroy,mhra}. In contrast to
RSA which is largely descriptive, such models yield spatially-explicit
predictions for patterns of animal space use
in the form of a probability density function (pdf),
by modeling the process of individual movement.
Mathematically, the
fine-scale behavior of individuals is treated as
a stochastic (Markov) process \cite{patlak,okubo,kareiva,turchin},
specifying the probability of an animal at a given
location moving to a subsequent location during a given time interval.
From this description one can
derive, in the limit of small time intervals, a
continuous-time partial differential equation (PDE) for
the evolution of the pdf.
For example, a recent analysis of coyote home ranges in
Yellowstone \cite{moorcroftroy} used a ``prey availability plus
conspecific avoidance'' (PA+CA) mechanistic home range model to account
for the observed patterns of coyote home ranges within the park.
In this model, individuals exhibited an avoidance response to
encounters with foreign scent marks and a foraging response to prey
availability (individuals decreased their mean step length in
response to increasing small mammal abundance in different habitats.)
In such mechanistic models,
inferences about long-term space use are usually made
by evolving the continuous-time PDE in order to converge to
its steady-state; this can be computationally
time-consuming \cite{moorcroftthesis,mhra}.

In recent work \cite{recon}, we developed a mechanistic home range
model which reconciles these two main approaches by combining the
concept of a spatial preference function $w(x)$ with a stochastic
model of fine-scale movement behavior. At each time-step the movement
of an individual is governed by its relative preference for the {\em local}
habitat
surrounding its current spatial position \ie the preference function
restricted to a region of size $L$ centered on the individual's
current location. The length scale $L$ has two roles: it is the
typical (jump) distance per time step, but also can be interpreted as
the distance over which the animal is able to {\em perceive}
differences in surrounding habitats. This new model, which we will call a
MEchanistic
RSA, or \melp\ model, is
compared against traditional RSA in Fig.~\ref{fig:bigpic}, and
detailed in Section \ref{sec:model}.
In \cite{recon} we were able to show that,
in one dimension (1D) with a spatially smooth habitat
preference function, using the Kramers-Moyal expansion
(\cite{stoc,recon} and App.\ A and E of \cite{mhra})
one may derive an an
advection-diffusion equation for the expected patterns of space
use, with advection and diffusion coefficients related
to the parameters of the underlying stochastic movement model. We then
showed that the resulting steady-state pdf could be determined
analytically, giving a intensity of space use proportional
to the square of the preference function.

In this paper we expand and generalize this result to the more
biologically relevant case of individuals moving in two space dimensions
across landscapes which may include discrete habitat types
giving rise to a {\em non-smooth} preference function.
Specifically, in Section \ref{sec:ex} we solve for
the steady-state pdf of our discrete-time
stochastic model of fine-scale movement behavior directly and
analytically, for arbitrary preference functions, without
resorting to the conventional procedure of taking the continuous-time limit
\footnote{%
One may ask whether discrete or continuous time is more appropriate
for animal movement modeling.
Clearly in reality animals move in continuous time, however a
continuous-time diffusion equation (Brownian process)
cannot be realistic on the shortest time-scales for the simple
reason that this would require infinite movement speed
(\eg see \cite{othmer}).
Therefore models with a fixed discrete time step remain crucial for
fine-scale modeling.}.
%
(Note that this relies on a particular algebraic feature of our model;
for a general redistribution kernel such an analytic solution
is not available.)
By doing so, we are able rapidly to compute exact steady-state
space use patterns for
the model for a full range of possible length scale
values $L$, rather than being confined to the limit of small $L$
inherent in a PDE-based approach.
We also gain an understanding of the numerically-observed
pdf behavior at jumps in preference function, which had remained
a mystery \cite{recon}.
A consequence of our solution
is that the computationally-intenstive task of solving
an inverse problem to fit multiple model parameters
can become orders of magnitude faster.
We discuss such numerical implications in Section~\ref{sec:num}, and
give CPU timings for our numerical examples throughout.

For illustrative purposes, we choose a
translationally-invariant exponential distribution of jump lengths, a
kernel which has proven useful for modeling coyote foraging movement
\cite{mhra}. However, our analytic solution also holds for a
generalization of the model of \cite{recon} to {\em spatially-varying}
diffusion coefficient (spatially-dependent length scale $L$) such as
occurs in modeling the prey-density dependent foraging rate of wolves
and coyotes \cite{white,lewis,moorcroftroy}.
In Section \ref{sec:ust}
we explore the transition from small $L$ (where the model tends to the
advection-diffusion equation derived in \cite{recon}), to large $L$
(where the model becomes equivalent to RSA). Here `large' and `small'
are in comparison to the typical spatial scale on which the preference
function changes.  In Section \ref{sec:ust} we explore this
numerically both in 1D, and with a 2D preference function derived from
discontinuous real-world small mammal abundance data appropriate for
coyotes.
In Section \ref{sec:evol} we show rapid and efficient numerical simulation
of the time evolution of the pdf.  Furthermore in Section
\ref{sec:jump} we analyse the behavior at a sharp discontinuity in
preference function, and show that on spatial scales larger than $L$
this effect may be approximated by effective matching conditions in
coupled advection-diffusion equations. Finally in Section
\ref{sec:conc} we discuss implications and draw conclusions.

\section{Mechanistic spatially-explicit (\protect\melp) model}
\label{sec:model}

The time-dependent pdf of an animal we will represent by $u(x,t)$,
where $x\in\Omega$ is the location, $\Omega\subset\mathbb{R}^d$
represents the habitat
region of interest, and $d$ is the dimensionality of space (usually 1 or 2).
Thus, in a 1D setting,
$u(x,t)dx$ is the probability that at time $t$ a given individual
is to be found in the interval $[x,x+dx]$.
(Note in 2D we use $x$ rather than $\mbf{x}$ to represent location vector.)
Its normalization is
\be
\ino u(x,t) dx \; = \; 1 \qquad \mbox{for all } t
\label{eq:nrm}
\ee
Our \melp\ model is an example of a
Markov process.
At each time step of length $\tau$
the current pdf at time $t$ is acted on by a fixed linear operator to
get the pdf at time $t+\tau$.
This is expressed by the master equation
\be
u(x,t+\tau) = \ino \ke(x,x') u(x',t) dx' \qquad \mbox{for } x\in\Omega
\label{eq:master}
\ee
Given an initial pdf $u(x,0)=u_0(x)$ for all $x\in\Omega$,
by iteration the pdf at arbitrarily large future times
(multiples of $\tau$) may be computed.
Here the redistribution operator kernel $\ke(x,x')$ is defined as
the conditional pdf of an individual animal's
location $x$, a time interval $\tau$ into the future, given
that its current location is $x'$
\footnote{Note that in the stochastic
literature the order of $x$ and $x'$ is often reversed \cite{stoc}}.
This is an uncorrelated jump or `kangaroo' process \cite{othmer}
(note the animal has no memory beyond the fixed time scale $\tau$).
Since a redistribution kernel is a conditional
pdf it is everywhere non-negative and (`columnwise') normalized by
\be
\ino \ke(x,x') dx = 1 \qquad \mbox{for all } x'\in\Omega
\label{eq:knrm}
\ee
Clearly for a given ecological situation the choice
of $\tau$ determines the form of $\ke(x,x')$.
For example, shorter $\tau$ may demand a smaller
kernel width $L$ simply because animal speed is limited.
By the length scale $L$ we mean the typical size of $|x-x'|$ for which
$\ke(x,x')$ is significant.
We will not indicate explicitly the dependence on $\tau$ of the form of
$\ke$.
The appropriate value of $\tau$
depends on the application; it needs to be large enough that
successive
animal relocations can be approximately treated as
uncorrelated.
Real-world location data collection technology also can be a factor
if fine-scale model fitting is to be done.
For coyotes, a typical value of $\tau$ is 10 minutes \cite{mhra}.


We consider a spatial preference (resource selection) function
$w\in L^1(\Omega)$
which controls relative preference for each location in the domain.
We represent unbiased (`preference-free') diffusive animal movement
with a {\em symmetric} redistribution kernel $\ph(x,x')$,
that is,
\be
\ph(x,x') = \ph(x',x) \qquad \mbox{for all } x,x'\in\Omega.
\ee
The kernel $\phi$ obeys the Markov normalization (\ref{eq:knrm});
from this and symmetry it follows that
\be
\ino \phi(x,x') dx' = 1 \qquad \mbox{for all } x\in\Omega
\label{eq:pinv}
\ee
which is the statement that a constant pdf
is invariant under redistribution by $\ph$.
Since a uniform density gives no net probability mass flow, we say
that (an operator with kernel) $\ph$
is {\em advection-free}.
Our \melp\ redistribution kernel $k$ is this advection-free
jump kernel biased by the preference function, in other words,
\be
\ke(x,x') = \frac{w(x)\ph(x,x')}{z(x')}
\label{eq:ke}
\ee
with normalization function (easily seen to be required to
satisfy (\ref{eq:knrm})),
\be
z(x'):=\ino w(x'')\ph(x'',x') dx''.
\label{eq:z}
\ee
For (\ref{eq:ke}) to be meaningful we must have $z>0$ everywhere;
it is sufficient that $w>0$ everywhere for this to hold,
which we will assume from now on.
%
Note that since $\ph(x,x')$ may depend independently on $x$ and $x'$
(barring the symmetry constraint), it may represent a
spatially-varying (and also anisotropic) diffusion coefficient.
The \melp\ model is thus more general than that of \cite{recon},
which was restricted to the translationally-invariant case
\be
\ph(x,x') = \phr(x-x').
\label{eq:trans}
\ee
Here 
$\phr(\cdot)$ is a function of relative displacement alone,
which limits one to a spatially-invariant distribution
of step-lengths in the underlying stochastic movement model.

\subsection{Limiting cases of the model} 
\label{sec:lim}

We now discuss two limiting forms of \melp.
Firstly,
consider the case $\ph(x,x') = 1/\mbox{vol}(\Omega)$ for all $x,x'\in\Omega$.
This corresponds to preference-free redistribution to a uniformly-random
location
in $\Omega$, without regard to current location. (\ref{eq:ke}) then becomes
\be
\ke(x,x') = \frac{w(x)}{\ino w(x'') dx''}.
\label{eq:llim}
\ee
Since this is independent of $x'$, within a {\em single} time step
(and for all future time steps) the master equation reaches its
steady-state pdf $u^* := \lim_{t\to\infty} u(\cdot,t)$
given by
\be
u^*(x) = C_1 w(x), \qquad \mbox{with normalization constant }\quad
C_1^{-1}=\ino w(x'') dx''.
\label{eq:rsa}
\ee
This is formally equivalent to a conventional (time-invariant) RSA model,
with linear dependence on preference.
This assertion is illustrated when
$\Omega$ is divided into regions $j=1\cdots m$ each of area $A_j$
and constant preference function $w_j$, for then (\ref{eq:rsa}) assumes the
more familiar RSA form $u_j = A_j w_j / (\sum_{i=1}^m A_k w_k)$
where $u_j$ is the probability of being in region $j$
\cite{johnsonrsa,manly,boycerev}, see Fig.~\ref{fig:bigpic}a.
In the case of an infinite domain such as $\Omega=\mathbb{R}^d$
(in which case $w$ alone delineates the habitat)
no constant normalizable $\ph$ exists; however, the above result
may be reproduced by considering the limit in which
a the kernel $\ph$ becomes much wider
than all spatial scales of interest
in the habitat ($L\to\infty$).
Then $\ph(x,x')$ tends to a constant for $x$ and $x'$ within the habitat,
and (\ref{eq:rsa}) again follows.
Thus for both cases above we will call this the $L\to\infty$ limit,
and state that in this limit our \melp\ model degenerates to RSA.

Secondly, consider the $L\to 0$ limit where $k$ tends to a diagonal kernel.
In order for time evolution to take place at all, we must
also take the limit $\tau\to0$. It is well known \cite{okubo,othmer}
that the correct way to balance
these two limits in order to reach a well-defined diffusion coefficient
is to choose the variance of the kernel $k$ to scale as $\tau$.
In the 1D case of smooth $w$,
and a translation-invariant kernel (\ref{eq:trans}),
we have derived \cite{recon}
that in this limit our \melp\ model gives a Fokker-Planck PDE with
known advection and diffusion coefficients.
From this we showed that the steady-state pdf is
\be
u^*(x) = C_2 w^2(x),
\qquad\mbox{with constant }\quad C_2^{-1} = \ino w^2(x'')dx'',
\label{eq:ustw2}
\ee
that is, quadratic in preference function.
(The integral is bounded since $w$ is smooth and in $L^1(\Omega)$).
We note that in this Fokker-Planck limit,
our model is equivalent to that of \cite{thermal}
with the `potential' function $U(x) = - 2D \log w(x)$ and constant
diffusion $d(x) = D$.

We will see in Section \ref{sec:ust} how these differing
$L\to\infty$ and $L\to0$ steady-state limits are reached in practice.

\section{Analytic formula for steady-state pdf}
\label{sec:ex}

The condition that $u^*$ be a steady-state pdf is that it be invariant
under the master equation (\ref{eq:master}), in other words,
\be
u^*(x) = \ino \ke(x,x') u^*(x') dx' \qquad \mbox{ for all } x\in\Omega.
\label{eq:ust}
\ee
Our main result is the following claim.
\begin{pro} 
A steady-state pdf for the model redistribution kernel (\ref{eq:ke})
is given by
\be
u^*(x) = C w(x) z(x), \qquad \mbox{ with constant } \quad
C^{-1} = \ino w(x') z(x') dx',
\label{eq:ex}
\ee
where the function $z$ is defined by (\ref{eq:z}).
\label{pro:ex}
\end{pro} 
The proposition is proved by substituting (\ref{eq:ke}) and
(\ref{eq:ex}) into the right side of
(\ref{eq:ust}) then noticing that $z$ cancels, allowing the simplification
\bea
\ino \ke(x,x') C w(x') z(x') dx' &=& C \ino w(x)\ph(x,x')w(x')dx'
\nonumber \\
&=&  C w(x)\ino\ph(x',x)w(x')dx'
\nonumber \\
&=&  C w(x)z(x),
\nonumber \\
&=&  u^*(x),
\label{eq:proof}
\eea
verifying (\ref{eq:ust}).
Crucially, it is the
assumption that $\ph$ is symmetric that allows us to proceed
from the first to second line.
Notice that once standard assumptions about ergodicity
are satisfied,
the steady-state $u^*$ is unique
(\eg see Doeblin condition in \cite{meyntweedie} p.396;
in our ecological context this is satisfied since 
we can
assume that there is always some randomness to animal motion, \ie
the kernel $\ph$ is
always somewhat spreading at each spatial location).
Since $w$ is assumed to be everywhere positive, so is $u^*$.

We now derive some secondary results on the
structure of $K$, giving intuition into the reason for
existence of the simple formula (\ref{eq:ex}).
They may be skipped if no further mathematical insight is required.
We switch to operator notation, expressing (\ref{eq:ust})
as $u^* = Ku^*$ where $K$ is a Markov operator.
Recall that a Markov operator is an integral operator with non-negative kernel
obeying (\ref{eq:knrm}),
which can be expressed $K^T 1 = 1$ where $1$ is the constant function
and $K^T$ the adjoint operator (with respect to uniform measure).
Our model (\ref{eq:ke}) expresses the factorization
\be
K=W\Phi Z^{-1},
\label{eq:fac}
\ee
where $\Phi$ is the (Markov)
operator defined by the integral kernel $\ph(x,x')$,
and $W$ and $Z$ are
the diagonal operators which multiply by the functions $w$ and $z$
respectively.
Thus the structure is an operator with constant invariant measure
sandwiched between two diagonal operators.
Note that $z = \Phi w$ since $\Phi$ is symmetric.
Then we have,
\begin{pro} 
With the assumption $w\ge c$ everywhere, for some $c>0$, our model Markov operator (\ref{eq:fac})
\ben
\item satisfies detailed balance,
that is,
\be
k(x,x')u^*(x') = k(x',x)u^*(x)\qquad \mbox{ for all }  x,x'\in\Omega,
\label{eq:detbal}
\ee
\item is self-adjoint with respect to the measure $1/u^*$,
and
\item has all eigenvalues real.
\een
\label{pro:sa}
\end{pro} 
The proof is as follows. We define $U^*$ to be the operator multiplying
by the function $u^*$; it can be written $U^* = ZW$. Using
(\ref{eq:fac}) gives $KU^* = W\Phi W$, explicitly symmetric.
In other words
\be
(KU^*)^T = KU^*,
\label{eq:kut}
\ee
which is equivalent to detailed balance (\ref{eq:detbal}).
Now we use $(\cdot,\cdot)$ to indicate the $L^2(\Omega)$ (real)
inner product with respect to uniform measure, and $(\cdot,\cdot)_{1/u^*}$
with respect to measure $1/u^*$.
Using (\ref{eq:kut}) and the boundedness of $1/u^*$ in the middle step we have
\be
(a, K b)_{1/u^*} = 
(\frac{a}{u^*},KU^* \frac{b}{u^*}) = (KU^* \frac{a}{u^*},\frac{b}{u^*})
= (Ka, b)_{1/u^*}\qquad \mbox{ for all } a,b\in
L^2(\Omega)
\ee
which proves part 2.
Part 3
immediately follows by self-adjointness.
We remark that the simple formula (\ref{eq:ex}) for $u^*$ can now
be seen as a result of the need to symmetrize the factorization (\ref{eq:fac}).

\subsection{Implications for fast numerical modeling}
\label{sec:num}

We discuss briefly why the above result is important in practice.
The analytic steady-state pdf formula (\ref{eq:ex}) is special because,
for a general redistribution kernel no such formula exists.
In particular, there is no formula for $u^*$ for commonly-used 
mechanistic kernels such as those expressed in polar coordinates with
exponential radial jumps
and a von Mises angular distribution (Ch. 3 of \cite{mhra}).
%
%
(In that work all steady-state distributions had to be computed in
the advection-diffusion limit,
often in a numerically-intensive fashion;
see App. G of \cite{mhra} and \cite{moorcroftthesis}).
In fitting model parameters to real-world location data, a large number of
such steady-state pdfs must be found
as part of the parameter-optimization process
(\eg Sec.~4.3 of \cite{mhra}).

Numerical solution of $u^*$ for any redistribution kernel
requires discretization of the domain into $N$ degrees of freedom.
In 2D the $N$ required for acceptable accuracy can be large (\eg $10^4$).
Solving for $u^*$ given a
general kernel is then an eigenvector problem involving
a (possibly dense) $N$-by-$N$ matrix discretization of that kernel.
The iterative solution of such large eigenproblems can be slow especially when
diffusion rates are small.
In constrast, the formula (\ref{eq:ex}) in our \melp\ model bypasses this
and requires only the computational effort of the
{\em single} matrix-vector multiplication
required to evaluate the (discretized) integral (\ref{eq:z}).
This is an acceleration by orders of magnitude.

There is a further numerical advantage to a special case of the \melp\ model.
Namely if $\ph$ is translationally-invariant
(diffusion coeffcient is spatially constant)
then the action of the (discretized)
convolution operator $\Phi$ may be computed in time $O(N \ln N)$
via the Fast Fourier Transform (FFT) \cite{numrec}, which for large $N$
is much faster than the $O(N^2)$ dense matrix-vector multiply.
This further speeds up computing $u^*$ via (\ref{eq:z}).
Finally, the time-evolution $u(x,t)$ can now be computed much
more efficiently.
A single time-step of the master equation (\ref{eq:master}) may be performed
with $O(N \ln N)$ effort by using (\ref{eq:fac})
in a {\em split-operator method}:
division by $z$, followed by FFT application of $\Phi$,
followed by multiplication by $w$.

We illustrate these numerical techniques and advantages below.
All CPU timings are reported using a single core of
a 2GHz Intel Core Duo processor running MATLAB 7.0 in GNU/Linux.

\begin{figure}[t]  
\bc\includegraphics[width=6in]{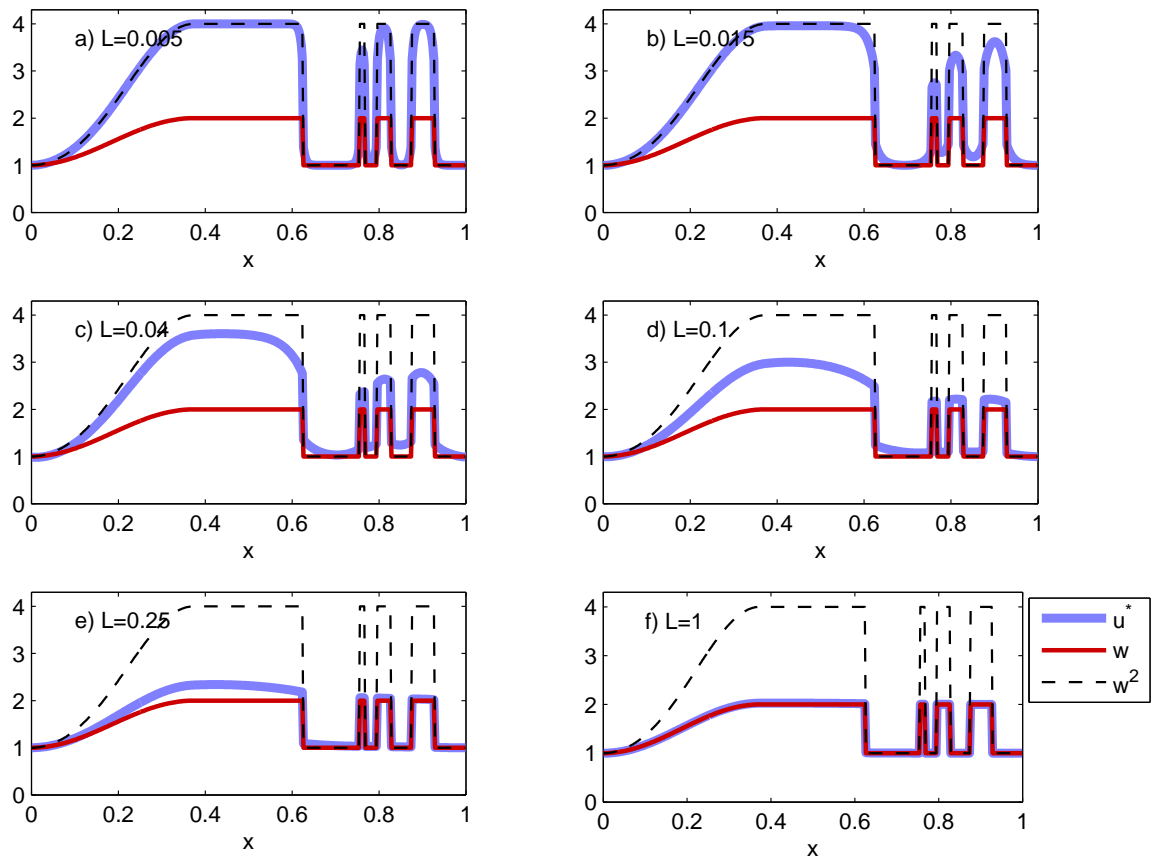}\ec
\caption{
\label{fig:phiseq}
Steady-state pdfs $u^*$ for exponential jump pdf
of (\ref{eq:exp}) for an increasing sequence of $L$ values.
The preference function $w$ (thin solid line) is chosen
to be smooth on the left side of the domain, and discontinuous
and oscillatory on the right side.
}
\end{figure} 

\begin{figure}[t] 
\bc
\includegraphics[width=3in]{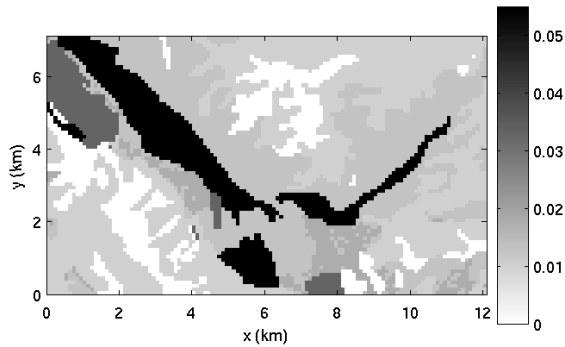}
\ec
\caption{
\label{fig:biomass}
Small-mammal abundance for Lamar Valley region of
Yellowstone National Park
collected by Crabtree \etal (unpublished); see Ch.~7 of \cite{mhra}.
Units are the total biomass density of small mammals in kg/ha with
darker colors indicating larger values.
The $x$- and $y$-axes are in kilometers.
The highest abundances of small mammals are found in the
mesic grassland habitats.
}
\end{figure} 

\begin{figure}[t] 
\bc
\includegraphics[width=\textwidth]{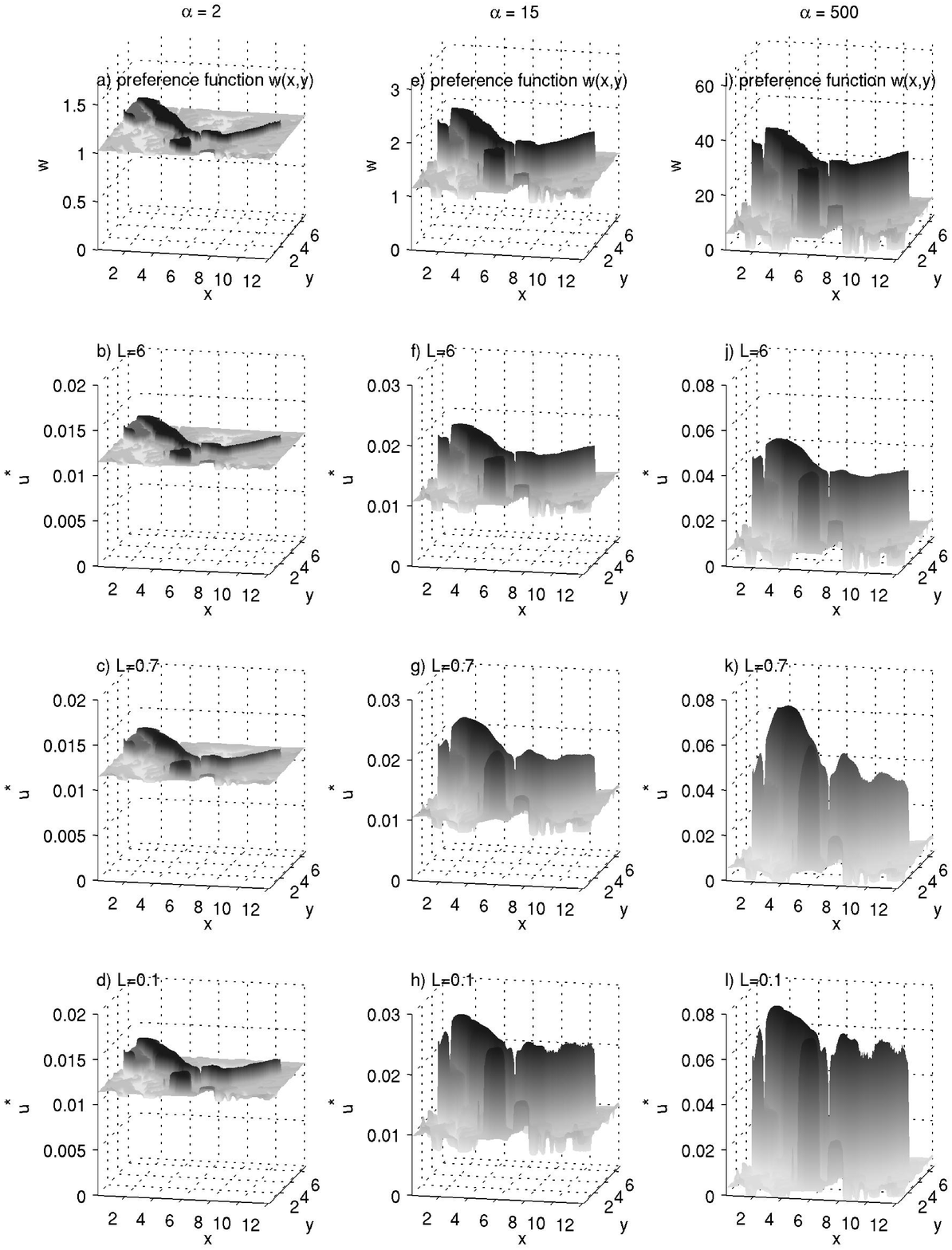}
\ec
\caption{
\label{fig:seq2d}
Steady-state 2D pdfs $u^*$ for the exponential jump kernel
(\ref{eq:exp2}),
applied to preference functions $w$
derived from biomass data shown in Fig.~\ref{fig:biomass}.
The three columns represent the choices $\alpha = 2, 15, 500$ in the
preference model (\ref{eq:biopref}).
At the top of each column is a surface plot of
the preference function, followed below
by surface plots of $u^*$ for three decreasing values of $L$.
}
\end{figure} 

\section{Numerical results for an exponential jump kernel}

In this section we illustrate the predictions of our \melp\ model
for steady-state and time evolution of space use for an idealized
1D preference function, and for a 2D preference function that is based on
spatially-complex, real-world measurements of prey abundance in different habitat types. 
Throughout we
consider a translationally-invariant $\ph$ kernel,
choosing an exponential kernel with width $L$, which takes the form in 1D
\be
\phr(\rho) = \frac{1}{2L}e^{-|\rho|/L}
\label{eq:exp}
\ee
where $\rho := x-x'$, and in 2D
\be
\phr(\rho) = \phi(|\rho|) =
\frac{1}{2\pi L}\frac{e^{-|\rho|/L}}{|\rho|}
\label{eq:exp2}
\ee
where we remind the reader that $\rho$ is a vector in the latter.
This 2D kernel has been used as a model of animal movement,
for instance describing the fine-scale movement of coyotes with remarkable
accuracy \cite{mhra}.
Note that the 2D kernel is radially symmetric, and when integrated
over angle it gives a distribution of jump distances $r:=|\rho|$
which is exponential, $(1/L)e^{-r/L}$.

\subsection{Steady-state}
\label{sec:ust}

Here we will assume the domain has periodic boundary conditions.
It has the interpretation that the piece of habitat $\Omega$ in question
is surrounded by similar (repeating) habitat, often a reasonable assumption.
%
%
Mathematically this condition is achieved in the case
of the unit interval $\Omega=[0,1)$ by replacing
(\ref{eq:exp}) by a sum over a few nearby `image' kernels,
\be
\phr(\rho) = \frac{1}{2L}\sum_{m=-M}^M e^{-|\rho-m|/L}
\label{eq:expimg}
\ee
where $M$ is chosen such that $\phr$ is periodic to some high
accuracy (\eg $10^{-6}$).
An analogous
2D image sum is used in the 2D case where $\Omega$ is a rectangle.

Note that our formalism can handle
other boundary conditions. For instance
we model a standard zero-flux boundary in Section \ref{sec:evol}.
We have also checked that the periodic steady-state
results are very similar to those
for zero-flux, except when $L$ is of order the size of the whole domain.
(In this case for zero-flux
$u^*$ is affected by the effective boundary discontinuity
in $w$, a complication which we will not pursue here).

\subsubsection{One-dimensional case}
\label{sec:ust1d}

In Fig.~\ref{fig:phiseq} we show $u^*$ computed via (\ref{eq:ex})
and (\ref{eq:z})
for the exponential kernel, for an ascending sequence of $L$ values,
in the unit interval.
Our example preference function $w$ has been chosen to exhibit a
variety of lengthscales: for $x<0.6$ it is smooth, corresponding to a
gradation in habitat preference, whereas for $x>0.6$ it is piecewise constant
with discontinuous oscillations between the values 1 and 2
corresponding to isolated patches of more favorable habitat.
The shortest length scale of $w$ is 0.015, namely the size of the smallest
constant patch near $x=0.76$.
We see that for very small $L$, $u^*$ accurately matches $w^2$ for
all regions apart from those with the most rapid $w$ variations.
This matches the expectation in Section \ref{sec:lim}.

A gradual transition is seen in the sequence of Fig.~\ref{fig:phiseq}.
As $L$ becomes larger than a given feature, $u^*$ in the vicinity
of that feature starts to become locally proportional to $w$.
Finally in plot (f), $L$ is larger than any feature and
$u^*$ globally becomes linear in $w$.
The explanation for this transition is simple and
lies with (\ref{eq:ex}) combined with the realisation that
the function $z$ is given by the function $w$ smoothed locally
over a width of about $L$. Fine ($<L$) features
in $w$ will thus be smoothed away giving a locally constant
$z$, whereas for coarse ($>L$) features $z \approx w$
and quadratic dependence in $w$ results.
This effect is common to any jump kernel with characteristic width $L$:
fine-scale habitat features result in $u^* \propto w$,
in accordance with the $L\to\infty$ limit (\ref{eq:rsa}),
whereas coarse-scale habitat features are tracked according
to $u^* \propto w^2$, in accordance with the Fokker-Planck
limit (\ref{eq:ustw2}).

Since $u^*$ is computed analytically via (\ref{eq:ex}),
its numerical accuracy is limited only by quadrature of the integral.
We computed $z$ using (\ref{eq:z}) via FFT convolution (Section \ref{sec:num})
in a few thousandths of a second on a uniform quadrature
grid of $N=400$ points.
This contrasts the order 1 s needed
to iteratively solve for
the dominant eigenvector of the dense matrix
discretization of $\Phi$ that would be required if no simplifying
factorization of $K$ or analytic solution for $u^*$ were known.

\subsubsection{Two-dimensional case with real-world habitat heterogeneity}
\label{sec:ust2d}

We start with the small-mammal density data $B(x)$
shown in Fig.~\ref{fig:biomass} where location $x$ was sampled on
a 0.1 km grid over a 7 km by 12 km domain $\Omega$.
This density is piecewise constant, being derived from
measured prey densities (mice, ground squirrels, 
pocket gophers and red-backed voles) appropriate for coyotes
in six different habitat types (see Ch.~7 of \cite{mhra}).
The main habitat feature is a strip of mesic grassland (the darkest
region in the figure), which follows a valley floor and
supports a high abundance of small mammals.

We use a linear relationship between prey availability and preference
\be
w(x) = 1 + \alpha B(x) \qquad \mbox{ for all } x \in \Omega
\label{eq:biopref}
\ee
where $\alpha$ (units of ha/kg) controls the strength of the
preference per unit of prey biomass. The resulting steady-state pdfs
$u^*$ are shown in Fig.~\ref{fig:seq2d}. This shows the patterns of
space use for 3 different values of $\alpha$, in combination with 3
different values of length scale $L$ (\ref{eq:exp2}). Each column
represents an $\alpha$ value, ranging from a weak preference (left
column) to strong (right column).  For comparison, the $w$ function is
shown at the top of each column.

The computational grid was moderately-sized ($N=8591$), the same
spatial resolution as the underlying estimates of prey abundance was
sampled. Once $\phi(\rho)$ had been evaluated on the grid, solving for
each steady-state pdf required only 0.009 s using (\ref{eq:ex}) with
$z$ computed from $w$ by 2D FFT convolution
\footnote{We have also tried other model preference functions on larger
grids: a 200 by 200 grid ($N=40000$) requires 0.025 to find $u^*$.
By contrast, note that solving for $u^*$ without the factorization
of $K$ is impractical
(even representing $K$ as a dense $N$-by-$N$
matrix on this grid would require 12 GB of memory).}%
.
This is 100-1000 times faster than an iterative solution
for the dominant eigenvector of $K$
if neither the factorization nor analytic formula are used
(a single dense matrix application of a general $K$ takes 0.45 s and many such
iterations are required for convergence, the number depending on $L$ and
the particular $w$).
Even if the factorization (split operator method) were used to perform
each iteration, our analytic formula (\ref{eq:ex}) would still be
10-100 times faster.

How is steady-state space use controlled by $\alpha$ and $L$?
Comparing the $L=6$ row (cases b,f,j) to the preference function itself
(a, e, i), we see that with this large length scale $u^*$ is very close
to proportional to $w$, as in 1D and as as explained
in Section \ref{sec:lim}.
Proceeding down the figure, we see smaller $L$ values result in a
$u^*$ with an exaggerated tendency towards space use becoming
increasingly concentrated in areas of higher preference. This results
in much more relative animal concentration in the mesic grassland
region than would be predicted by traditional RSA. This tendency has
reached its limit by the bottom $L=0.1$ row (d, h, l), where $u^*$ is
close to proportional to $w^2$.

Consider the right-hand column (j, k, l).
Changing from $L=6$ to $L=0.7$ causes a substantial increase in relative
space use in the western part of the mesic grassland
(see the large bump to the left in k), but very little change in the
eastern part of this same habitat type.
The explanation is simple: the grassland strip is generally wider
than $L=0.7$ on the western side resulting in
a local tendency towards $u^*\propto w^2$,
but narrower than $L=0.7$ on the eastern side giving here
$u^*\propto w$.
Finally in case l $L=0.1$ is narrower than all parts of the mesic
grassland and the densities equalize on west and east sides. 
This is analogous to the transition discussed in Section \ref{sec:ust1d}.
This interesting geometric effect is absent in traditional RSA.

\begin{figure}[th] 
a)\quad \raisebox{-1.6in}{\includegraphics[width=2.5in]{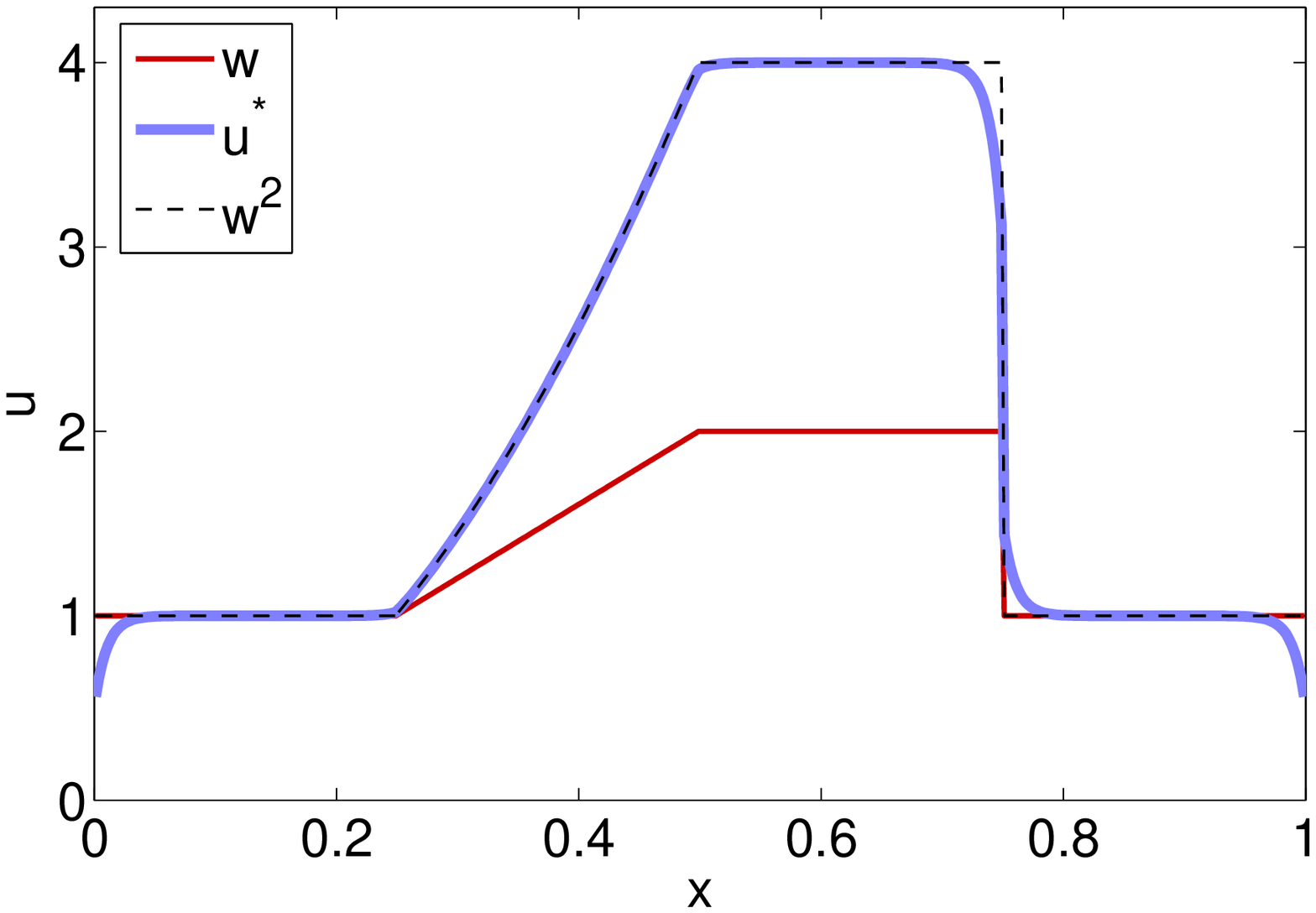}}

\includegraphics[width=5in]{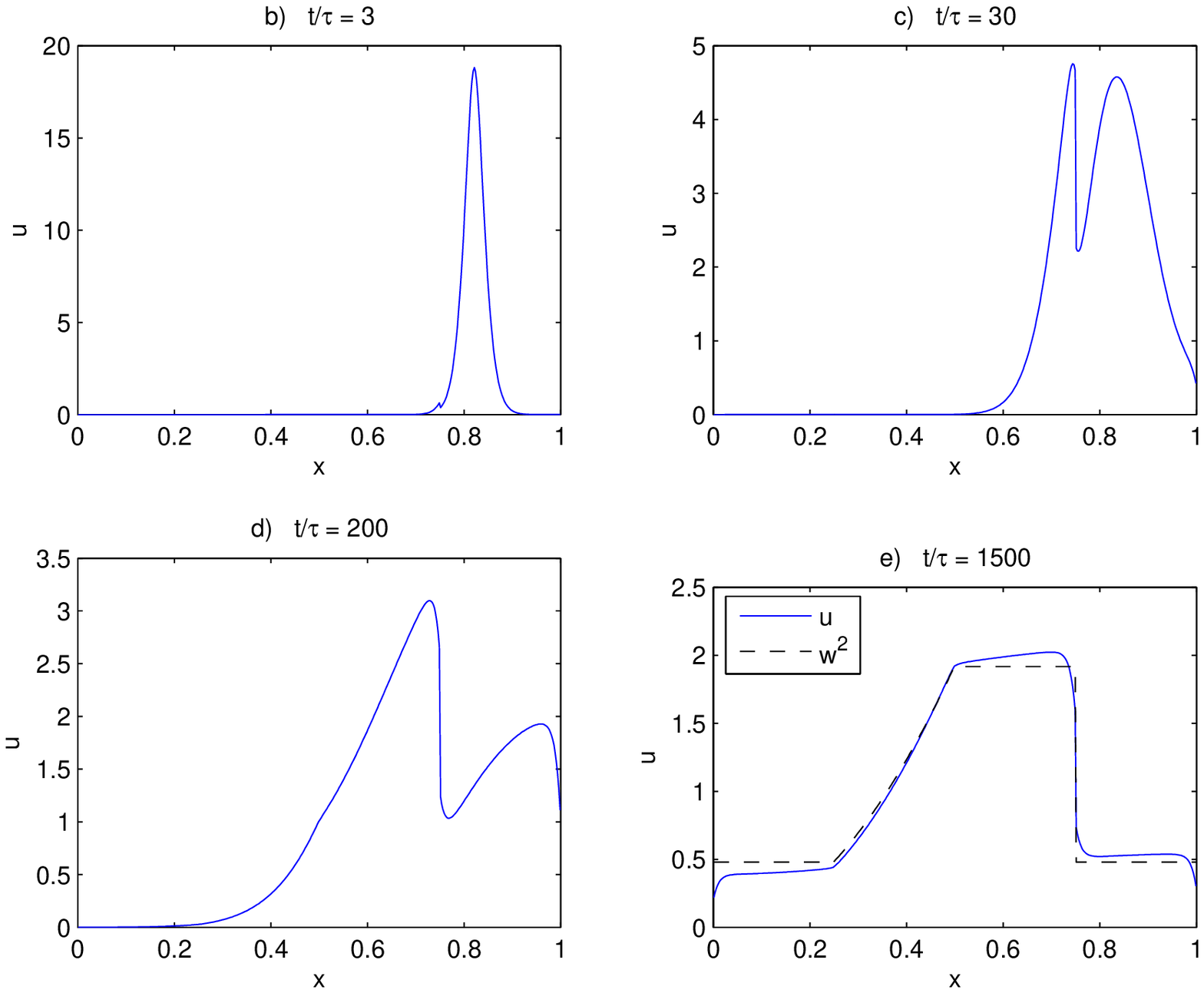}
\caption{
\label{fig:evol}
a) Piecewise linear model preference function $w(x)$, steady state pdf
$u^*(x)$,
and predicted limiting case of steady-state pdf (\ref{eq:ustw2}) (dashed).
b)-e) Snapshots of
time evolution of $u(x,t)$ under master equation (\ref{eq:master}) in 1D,
at four times (indicated by $t/\tau$ the number of iterations).
The exponential jump pdf of (\ref{eq:exp}) is used with $L=0.01$.
In e) $u$ is very close to steady-state; the prediction
(\ref{eq:ustw2}) is also shown (dashed).
}
\end{figure} 

\begin{figure}[th] 
\bc
\includegraphics[width=6in]{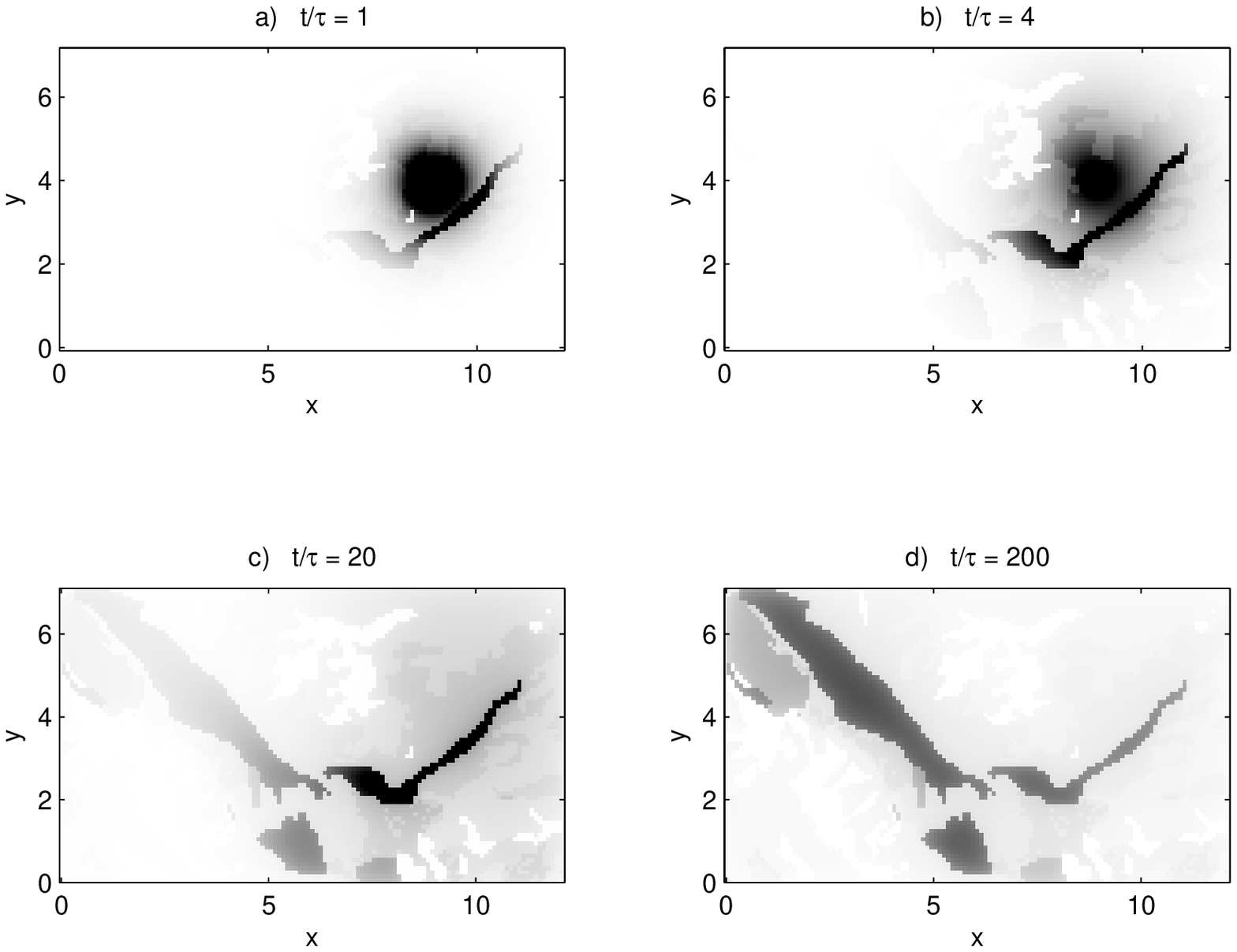}
\ec
\caption{
\label{fig:evol2d}
Density plots snapshots
of the time evolution of $u(x,t)$ under master equation
(\ref{eq:master}) in 2D, at four times
(indicated by $t/\tau$ the number of iterations).
The exponential jump pdf of (\ref{eq:exp2}) is used with $L=0.7$,
and preference function
derived from biomass data in Fig.~\ref{fig:biomass} via Eq.(\ref{eq:biopref})
with $\alpha=500$.
In d) $u$ is very close to steady-state,
which is shown in Fig.~\ref{fig:seq2d}k.
}
\end{figure} 

\subsection{Evolution in time}
\label{sec:evol}

In Figure ~\ref{fig:evol} we show the time-dependent evolution of
space use $u(x,t)$ under the master equation (\ref{eq:master}) in a
single space dimension for simple piecewise linear model preference
function with a single discontinuity (see panel a), with exponential
kernel with length scale $L=0.01$.
Zero-flux boundary
conditions were created by using the non-periodic jump kernel
(\ref{eq:exp}). Panels b-e show the resulting dynamics of space use,
starting from an initial condition $u_0(x) = \delta(x-x_0)$ where
$x_0=0.82$
%

The computation with $N=400$ grid points took 0.0005 s per time step;
this was done with iterated multiplication by the dense $K$ matrix
since the non-periodic choice of $\phr$ makes the $\Phi$ operator
no longer a convolution. Notice the discontinuity feature in $u$ at $x=0.75$ establishes itself rapidly and persists throughout
the full time range.

Fig.~\ref{fig:evol2d} shows the time-depednent evolution of space use
for the 2D real-world biomass preference model (\ref{eq:biopref}) with
$\alpha=500$ and $L=0.7$ (same parameters as Fig.~\ref{fig:seq2d}k),
for a delta-function initial condition at $x_0 = (9,4)$.  After a
single iteration (panel a), the local preference biases the individual's
movements towards the narrow nearby eastern strip of mesic
grassland. In panel (b) its space use is still split between an
expanding radial distribution about its initial position and the
nearby grassland. In panel (c), the compounded mechanistic movement
steps have caused the individual's space use to become concentrated in
the eastern strip of mesic grassland but the mesic grasslands in the
western portion of the landscape are, at this stage, mostly
unoccupied. Much later in the simulation however (panel d), the intensity
of space use in the western mesic grassland areas is higher than the
eastern mesic grasslands, due to the geometric effects on the steady-state
$u^*$ described in Section \ref{sec:ust2d}.

The simulation from which the snapshots
in Figure \ref{fig:evol2d} were extracted is very rapid,
animating smoothly in real time at 30 frames/s even though $N \approx
10^4$, allowing immediate interactive model exploration. The raw
calculation (no graphical animation) takes 0.008 s per time-step,
benefitting greatly from the split operator method using FFT
convolution. Performing this without the aid of the factorization
(\ref{eq:fac}) is about 100 times slower. Furthermore, we find that the
additional effort needed to extract the mean-square displacement
$<x^2>(t) := \ino (x-x_0)^2 u(x,t) dx$,
a useful measure of spreading,
is negligible. We remark that
in order to model zero-flux boundary conditions while still using FFT
convolution, a periodized kernel was used but the $w$ array was
zero-padded with a border of width $O(L)$, at negligible extra cost.
%

\begin{figure}[t] 
\bc
a)\quad\raisebox{-2in}{\includegraphics[width=2.8in]{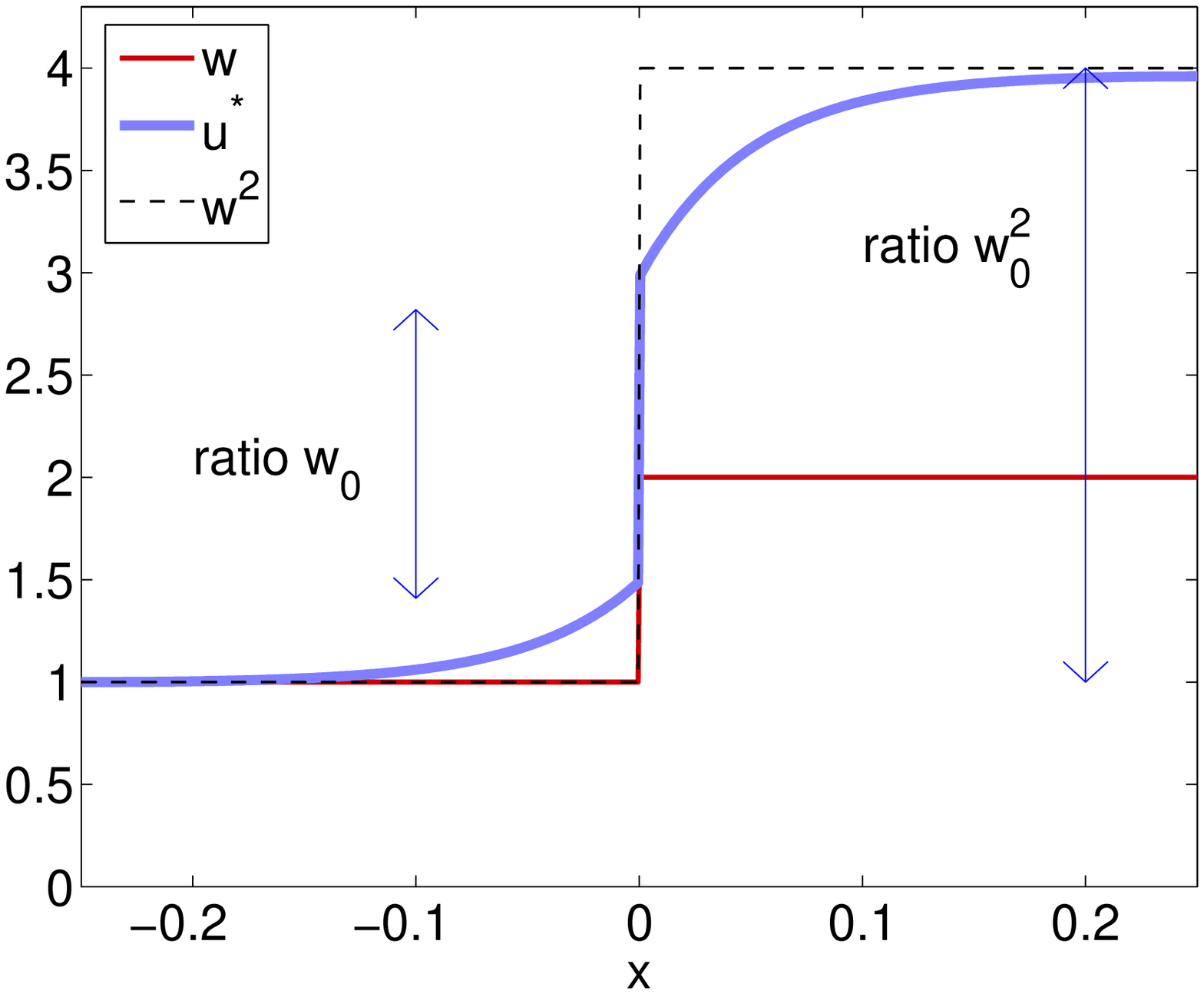}}\hfill
b)\quad\raisebox{-2in}{\includegraphics[width=2.8in]{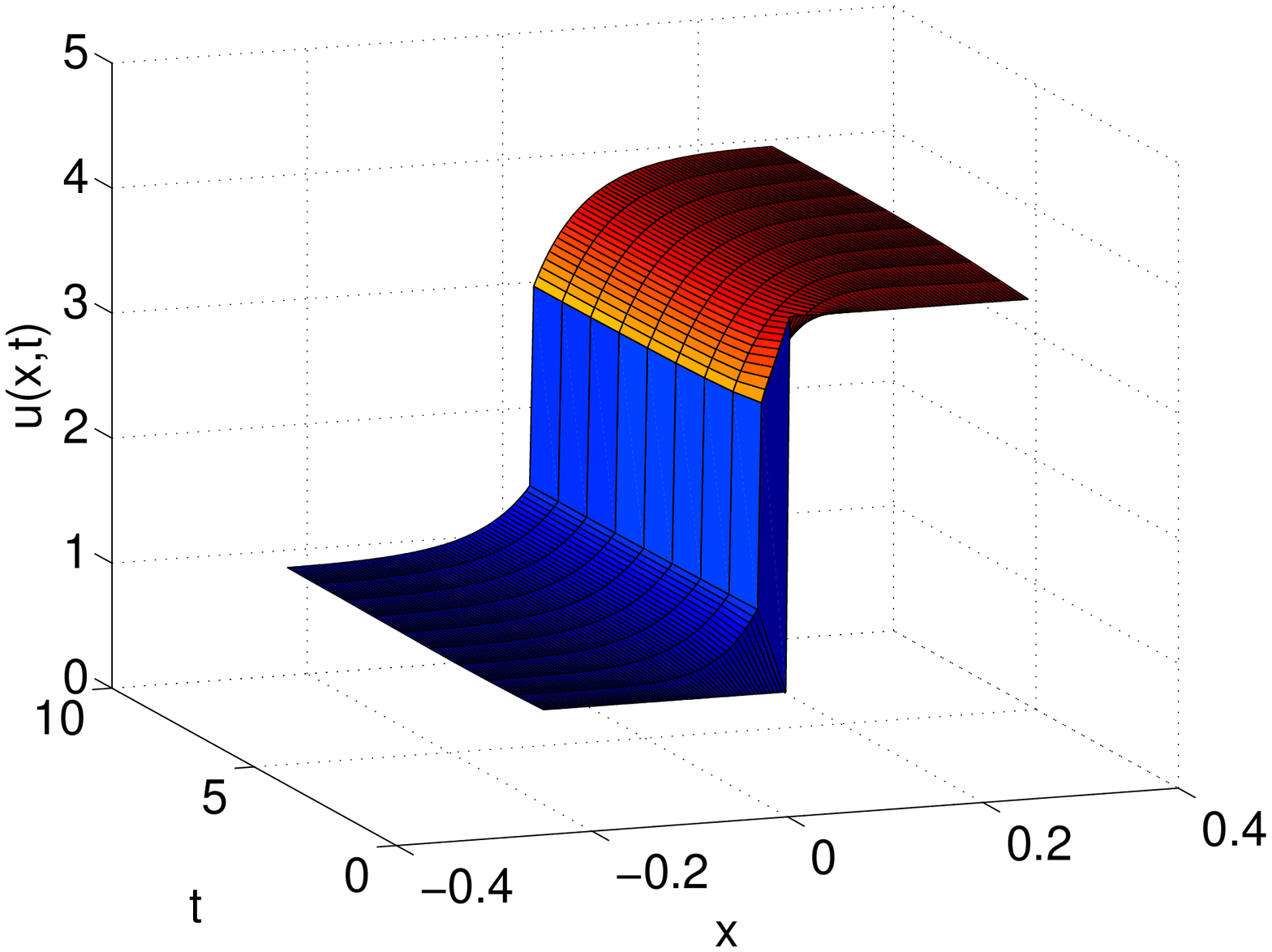}}
\ec
\caption{
\label{fig:expjump}
a) Detail of transition in steady-state space use
occurring within distance $L$ of 
a discontinuity
in preference function of the form (\ref{eq:wo}) with
$w_0=2$, for the 1D exponential jump kernel
of (\ref{eq:exp}). 
b) Time-evolution of $u(x,t)$ in this same local region for 10 time-steps
of the master equation (\ref{eq:master}), starting from an initial
pdf $u_0(x) = w(x)^2$.
}
\end{figure} 

\section{Local behavior near a sharp habitat transition}
\label{sec:jump}

We now examine in more detail what takes place at the discontinuities
in preference function $w(x)$ that arise at the boundaries between
different habitat types. As noted earlier, such discontinuities were
not able to be treated in the derivation of the advection-diffusion
limit in previous work \cite{recon}.

For simplicity, we consider a single discontinunity on a 1D landscape,
however we expect, and observe, similar behavior in the more
biologically relevant case of multiple discontinuities arising at the
edges of different habitat types on a 2D landscape. Consider a
landscape in which there exists a single boundary between two habitat
types located at $x=0$, resulting in the following discontinous
preference function:
\be
w(x) = \left\{\begin{array}{ll}1,& x<0\\w_0, & x>0\end{array}\right.
\label{eq:wo}
\ee
Given a jump kernel of width $L$ it follows from (\ref{eq:z}) that
$z(x)$ is a smoothed
(mollified) step-function with transition region width $L$.
The analytic formula (\ref{eq:ex}) then tells us that the steady-state pdf
jumps by a ratio of $w_0$ precisely at the habitat transition;
however, 
when viewed on length scales larger than $L$, the pdf jumps
by a ratio $w_0^2$. For the 1D exponential kernel (\ref{eq:exp}),
the analytic expression for the steady-state 
follows from that of $z$ via (\ref{eq:ke}), and is
\be
u^*(x) = \left\{\begin{array}{ll} 1 + \frac{W-1}{2} e^{x/L},& x<0\\
W\left[W - \frac{W-1}{2}e^{-x/L}\right], & x>0\end{array}\right.
\ee
The spatial decay length is thus the same as that of the kernel $\ph$.
This is shown in Fig.~\ref{fig:expjump}a.

How fast is this equilibrium reached? We
demonstrate in Fig.~\ref{fig:expjump}b that
the transition region's shape reaches its approximate
equilibrium in only a single time-step
(a slight change also occurs over the next few time-steps).
Here we chose initial conditions which already matched the global
pdf ratio of $w_0^2$, in order to study the equilibration in this
local region alone.

\subsection{Effective boundary condition for advection-diffusion equation}

Armed with this understanding of the local behavior at a
preference function discontinuity, we can incorporate
this into the Fokker-Planck PDE model
for the evolution of $u(x,t)$ in the
$\tau\to 0$ limit (in which case kernel width $L$ must also go to zero at an
appropriate rate).
We remind the reader that the Fokker-Planck equation is
\be
\pd{t}u = - \pd{x}[ c(x) u ] + \pdd{x}[ d(x) u]
\ee
where $c(x)$ and $d(x)$, representing drift and diffusion rates,
take on values given by the $\tau$-scaling of the second moment
of the jump kernel $\ph$ \cite{stoc,mhra,recon}

We combine two observations: i) locally the steady-state in the vicinity
of a discontinuity in $w$ enforces a multiplicative
jump (the square of the $w$ ratio) in $u$, and ii)
in the $\tau\to0$ limit the Fokker-Planck equation evolves on much
slower time-scales than the local equilibration in this vicinity
(which happens in $O(\tau)$).
Thus we expect that, for evolution on time-scales long relative to $\tau$
the transition region is in {\em local equilibrium}, with the
effective boundary condition
\be
\frac{u(x_-,t)}{(w_-)^2} =  \frac{u(x_+,t)}{(w_+)^2},
\qquad \mbox{for all } t>0,
\label{eq:effbc}
\ee
where the subscripts $-$ and $+$ indicate limiting values on the
left and right side of the discontinuity respectively.
Similarly, by conservation of the flux $J(x) = \pd{x}[d(x) u] - c(x) u$
across the discontinuity, we must have that if $d(x)$ is continuous
and $c(x)=0$ then 
\be
\pd{x}u(x_-,t) =  \pd{x}u(x_+,t), \qquad \mbox{for all } t>0,
\label{eq:effbcd}
\ee
at a step discontinuity in $w$.
We may now interpret (\ref{eq:effbc}) and (\ref{eq:effbcd})
as {\em matching conditions} for coupled
advection-diffusion equations on either side of the discontinuity.
In this way we have a recipe to understand the diffusion limit
even in the presence of discontinuous preference functions.
We remark that our assumption of locally vanishing $c$ corresponds
to no gradient in $w$ locally on either side
(the case with general values of $w'$ either side we postpone for future work).
The above argument is non-rigorous, relying on reasoning
based on separation of length- and time-scales.
However it seems to be supported qualitatively by the evidence in
Fig.~\ref{fig:evol}b-e (although at early times (\ref{eq:effbc}) does
not appear to hold accurately).
We suggest that a more detailed analysis via
matched asymptotic expansions should be carried out.

\section{Conclusions}
\label{sec:conc}

We have analyzed the mathematical properties of a mechanistic resource
selection
(\melp)
model that captures the influence of spatially-localized
habitat preference on the movement behavior of individuals and
predicts their resulting patterns of space use.
The model combines random foraging motion with a local sensitivity
to habitat preference over a `perceptual radius' $L$.
Directed
movement bias is generated
in a similar manner to the angle-biased (von Mises) jump
kernels used in \cite{mhra}, and also becomes equivalent in the
small-$L$ limit to continuous-time advection-diffusion in a `confining
potential' as in \cite{thermal}.
Our analysis shows that the model has a desirable
factorization (\ref{eq:fac})
which yields a simple
closed-form formula (\ref{eq:ex}) for the steady state pdf.
The effect of the compounded random
movement decisions upon this steady state pdf
is a novel geometry-dependent scaling
with the preference function: linear when $L$ is large
(compared to local habitat features), but quadratic when $L$ is small.

These novel spatial effects are absent in conventional RSA, and have
only been analysed previously in mechanistic home range models in the
context of the advection-diffusion limit \cite{recon} in which the
perceptual radius of individuals is small ($L \to 0$). Our analytic
formula developed here allows this to be understood for the case of
discontinuous preference functions and for any given perceptual radius
$L$.

Large gains in
computational efficiency have been demonstrated throughout, including
the case of discontinuous 2D preference functions motivated by
observations of spatially-varying prey availability across different
habitat types. We believe such efficient forward models will be
important tools, as inverse modeling, and fitting of multiple model
parameters to observations, become more popular and time-consuming.

Although we did not construct or study them numerically in this work,
we expect that the benefits demonstrated here for a 
model that is linear in $u$
(with prey- or spatially-
dependent movement rates) will also apply in the analysis of {\em
non-linear} models such as those with density-dependent diffusion (\eg
Sec.~3 of \cite{white}), and
scent-mediated interactions between multiple
animal packs \cite{white,lewis,mhra}.
For example, chemotaxis could
be included in the preference function $w$, in which case our
analytic formula
(\ref{eq:ex}) could be used to turn a coupled PDE system into coupled
algebraic systems, a huge simplification.
We also expect that by extending our preliminary operator analysis
(Proposition \ref{pro:sa}), the spectral properties,
and hence equilibration rates of animal home range
space use, may be deduced.


Concise Matlab codes for computation of all figures in this work are
freely available at {\tt http://math.dartmouth.edu/$\sim$ahb/moorcroft/}

\acknowledgments
AHB and PRM thank Robert L. Crabtree for the dataset on small mammal
abundance in Yellowstone National Park.
AHB thanks Jonathan Goodman and Luc Rey-Bellet for useful discussions
on the properties of Markov operators.
AHB is partially funded by NSF-0507614.



\bibliographystyle{abbrv}    
\bibliography{alex}

\end{document}